# Conformal Mapping for Multiple Terminals

Weimin Wang[1,*], Wenying Ma[2], Qiang Wang[1], Hao Ren[3]

[1]*State Key Laboratory of Optical Technologies on Nano-Fabrication and Micro-Engineering, Institute of Optics and Electronics, Chinese Academy of Science, Chengdu 610209, China*
[2]*College of Communication Engineering, Chengdu University of Information Technology, Chengdu 610225, China*
[3]*School of Electrical, Computer, and Energy Engineering, Arizona State University, Tempe, Arizona 85281, USA*

**Abstract**
Conformal mapping is an important mathematical tool in many physical and engineering fields, especially in electrostatics, fluid mechanics, classical mechanics, and transformation optics. However in the existing textbooks and literatures, it is only adopted to solve the problems which have only two terminals. Two terminals with electric potential differences, pressure difference, optical path difference, etc., can be mapped conformally onto a solvable structure, e.g., a rectangle, where the two terminals are mapped onto two opposite edges of the rectangle. Here we show a conformal mapping method for multiple terminals, which is more common in practical applications. Through accurate analysis of the boundary conditions, additional terminals or boundaries are folded in the inner of the mapped rectangle. Then the solution will not be influenced. The method is described in several typical situations and two application examples are detailed. The first example is an electrostatic actuator with three electrodes. A previous literature dealt with this problem by approximately treat the three electrodes as two electrodes. Based on the proposed method, a preciser result is achieved in our paper. The second example is a light beam splitter designed by transformation optics, which is recently attracting growing interests around the world. The splitter has three ports, one for input and two for output. Based on the proposed method, a relatively simple and precise solution compared with previously reported results is obtained.

---
[*] Correspondence to weiminwang@qq.com

## Introduction

Conformal mapping, as one of the most powerful tools of complex analysis, have been applied in many mathematical and physical fields. It mapped the unknown region onto a solvable structure, is especially suitable for two-dimension problems, such as transmission lines [1-5], integrated circuit components [6-11], electrostatic actuators [12-16], transformation optics [17-21].

Fig. 1a shows a representative example of electrostatics. The solid lines on the closed curve represent the surface of conductors and they are equipotential lines. The dashed lines represent electric field lines. Through a conformal mapping, the interior of the closed curve, i.e., the shaded region, is mapped onto a perfect (fringe effect-free) parallel-plate capacitor (Fig. 1b), whose electric field distribution is well known. Then the capacitance, potential, field, and charge of the unknown region can be obtained through an inverse mapping. The same method can also be utilized to magnetostatic field, all types of flow problems, transformation optics, and so on. Besides parallel-plate capacitor, other structures with known potential distribution, e.g., Figs 1c and 1d, can also be used as the mapped objects.

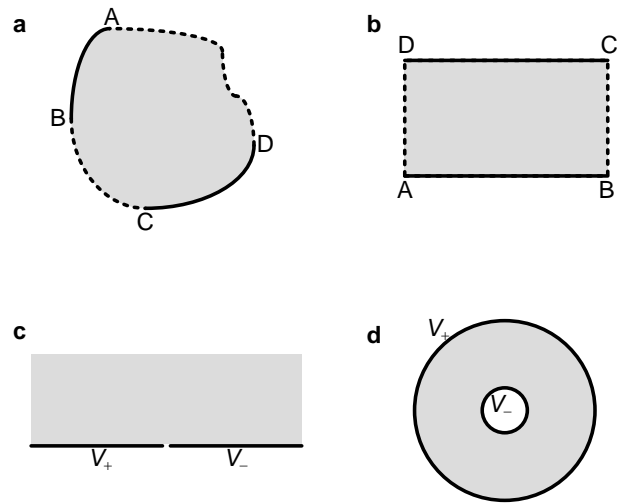

**Figure 1 | A representative example of conformal mapping. a,** The shaded region is a unknown structure. The solid lines are conductors and the dashed lines are electric field lines. **b,** The unknown structure is mapped conformally onto a perfect parallel-plate capacitor. **c,** Another structure with known potential and field distribution besides parallel-plate capacitor. The negative real axis has a potential of $V_+$ and the positive real axis has another potential of $V_-$. Thus the concentric circles whose center is origin are electric field lines and the straight lines through the origin are equipotential lines. **d,** The third structure with known potential and field distribution. Its electric field lines and equipotential lines are just opposite to those of **c**.

It is easy to see that this method is only suitable for two terminals with electric potential difference, pressure difference, optical path difference, etc. For more than two terminals, it is common to reduce the structure to two terminals by symmetry considerations. However, in many practical problems the structures are complicated and asymmetric. A general way is to approximately treat some terminals as one terminal and the other terminals as another [16, 22, 23]. If the potential of the multiple terminals in the former or the latter are different, they are approximately treated as the same firstly [14, 15, 24]. Then the problem is simplified to two terminals.

In this study a strict and exact conformal mapping for multiple terminals with arbitrary potential or net charge is proposed and described. Based on this proposed method, two application examples are detailed and compared with previous literatures.

**Mapping method**

The method is described in electrostatic problem, where the terminals are the surface of conductors. Taking three conductors as an example, as indicated in Fig. 2a, there are two situations for the extra conductor EF, i.e., a given potential or a given net free charge. First we consider that the conductor EF has the same potential with one of the other two conductors. Without loss of generality, we assume the conductors EF and CD have the potential $V$ and the conductor AB is grounded. It is easy to see that the directions of electric field lines BC and FA are from high potential to low potential, as shown as the solid arrows in Fig. 2a. On the other hand, due to the fact that the vertices D and E have the same potential, the direction of the electric field line DE is as the hollow arrows shown in Fig. 2a. In other words, from vertex D to vertex E along the dashed line DE, the potential first decreases and then increases. Therefore, a point D' exists at which the minimum potential occurs in line DE. As DE is an electric field line, the mapped line of DE in the perfect parallel-plate capacitor should be perpendicular to the mapped line of the conductors. Based on this analysis, the mapped capacitor of Fig. 2a is shown in Fig. 2b. The position of D' can be achieved through the relationship that the conductors CD and EF have the same potential, i.e., the length of DD' should be equal to the length of D'E. After the mapping is constructed, the electric field and potential in this folded rectangle is the same with a conventional rectangle. Then the original problem is solved.

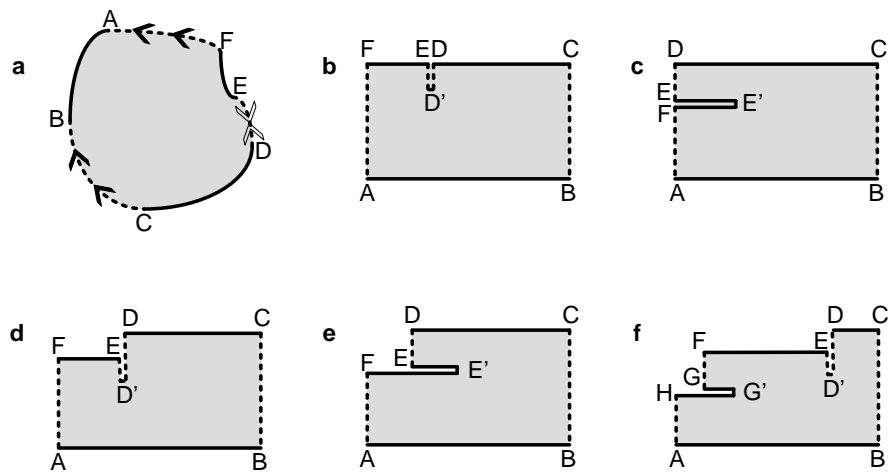

**Figure 2 | The conformal mapping for multiple conductors. a**, The extra conductor EF has the same potential with CD. Thus the directions of the electric field lines BC, DE, and FA are as the arrows shown. **b**, The mapped perfect capacitor of **a**. The dashed lines DD' and D'E overlap, although they are drawn separately to show that the structure is closed. **c**, The mapped capacitor of **a** when the conductor EF is floating and has a zero net charge. **d**, A possible mapped result of the situation that the potential of the conductor EF is different from the other two conductors. **e**, Another possible mapped result of the

situation **d**. **f**, An example of the conformal mapping for four conductors.

Second we consider that the conductor EF is floating and has a zero net charge. Now the conductor EF is an equipotential line and its potential is unknown. The equipotential line is parallel to the two conductors in a parallel-plate capacitor. Thus the corresponding mapped structure is shown in Fig. 2c. Given that the charge of EE' has the opposite polarity to the charge of E'F, and they have the same area density of charge, the length of EE' should be equal to the length of E'F. Therefore, the position of E' and the potential of EF can be solved through this relationship. Then this situation is also solved.

Next let us consider a more complex situation, e.g., the potential of the conductor EF is different from that of the other two conductors. For example, the potential of EF is smaller than that of CD and larger than that of AB. Under different structural parameters, the unknown region can be mapped onto Fig. 2d or 2e. Because the ratio of the potential of EF to the potential of CD equals to the ratio of the length of FA to the length of BC, the position of D' (Fig. 2d) or E' (Fid. 2e) can be solved, and which one of Figs 2d and 2e is the right mapped pattern will be determined. Then the mapped structure can be calculated in terms of two capacitors in parallel connection.

As for the situation that the conductor EF is floating and has a non-zero net charge $Q$, it is similar to the above situation, only that the net charge is known and the potential is unknown. Therefore the mapped pattern is also Fig. 2d or 2e. Given that all conductors have the same and uniform area density of charge in a perfect parallel-plate capacitor, the position of D' or E' can be determined by the following equations

$$\begin{cases} \varepsilon V \dfrac{L_{EF}}{L_{BC}} = Q & \text{for Fig. 2d} \\ \varepsilon V \dfrac{(L_{EF} - L_{EE'})}{L_{BC}} = Q & \text{for Fig. 2e} \end{cases} \quad (1)$$

where $V$ is the potential of CD, $\varepsilon$ is the dielectric constant of the shaded region and $L$ is the length between corresponding vertices.

All possible situations for three conductors have been discussed. If there are more conductors, obviously more edges will be folded in the inner of the mapped rectangle. One of the mapped results for four conductors is shown in Fig. 2f. It should be noted that here the positions of D' and G' are general determined by a system of quadratic equations with two unknowns.

Above we only discussed that the region to be solved is mapped onto a rectangle. Actually, due to the Riemann mapping theorem [25], any simply connected region, e.g., Fig. 1c, can be mapped conformally onto a rectangle. As for multiply connected region, a general structure is shown in Fig. 3a. Assuming that $V_-<V<V_+$, Fig. 3a can always be mapped conformally onto a circular-slit region, as shown in Fig. 3b. Then it can be further mapped onto a rectangle through a logarithmic mapping.

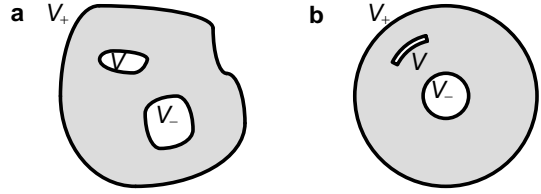

**Figure 3 | The conformal mapping for multiply connected region. a**, Two conductors with the potential of $V$ and $V_-$ are enclosed by the conductor with a potential of $V_+$. **b**, The corresponding conformal mapping of **a**, where the conductor with potential $V$ is an infinitely thin closed arc, whose center is the same with that of the conductor $V_-$ and $V_+$.

**Application examples**

Ref. [16] dealt with an array of electrostatic repulsive actuator, as shown in Fig. 4a. BC, DE and FGH are three conductors and their potentials are 0, $V$, $V$, respectively. The conductors BC and DE are fixed by an external force and the conductor FGH can only move in the direction of $y$-axis. Because of the existence of BC and DE, FGH is driven to move vertically upward. The electrostatic force applied on FGH can be obtained by calculating the change of electrostatic energy for different lengths of EF. He *et al.* approximately connected DE and FGH at infinity to reduce one conductor, as indicated in Fig. 4b. They both calculated theoretically the electrostatic force using conformal mapping and simulated it via Electro software. The structural parameters they used is listed in Fig. 5 of the literature. Their results are shown as the solid line and dashed line respectively in Fig. 5a. When the distance between the conductors DE and FGH is small, the results of conformal mapping and the simulation results coincide. However, the difference between them grows as the distance increases.

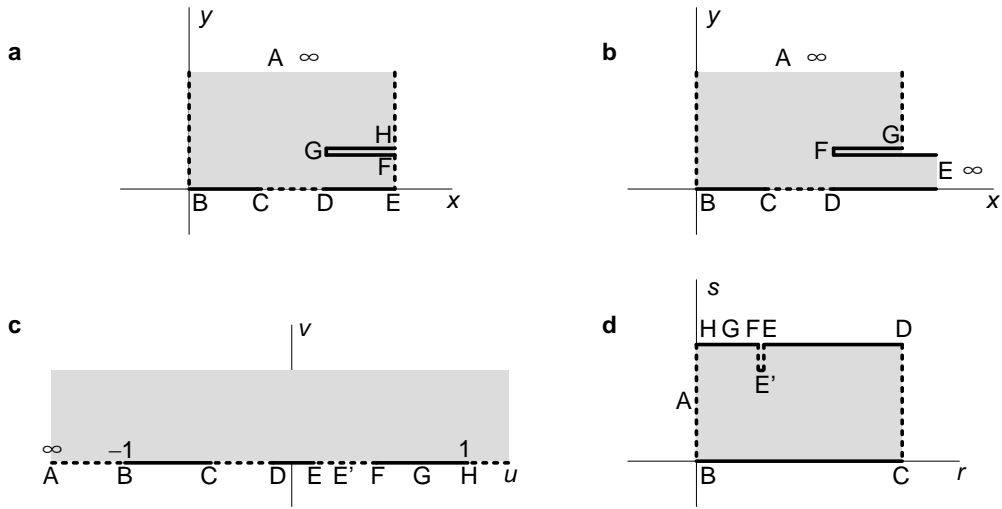

**Figure 4 | The problem to be solved in Ref. [16]. a**, The modelled structure of the problem. **b**, The simplified structure in Ref. [16]. **c**, The structure **a** is mapped conformally onto the upper half plane. **d**, **c** is further mapped onto a parallel-plate capacitor.

First we repeat the calculation of Fig. 4b of the literature and the results are indicated by the circular dots in Fig. 5a. They closely match the theoretical results of [16]. Next we strictly calculate the actuator, Fig. 4a. It is a generalized polygon. A complex plane, $z$-plane, is constructed such that vertex B is the origin and BE is the real axis. Thus, the coordinates of all vertices can be derived in $z$-plane based on the structural parameters. Through an inverse Schwarz-Christoffel (SC) mapping, Fig. 4a can be mapped onto the upper half-plane of another complex plane, namely, $w$-plane, as shown in Fig. 4c. The coordinates of any three vertices in $w$-plane can be chosen arbitrarily, e.g., −1 (vertex B), 1 (vertex H) and infinity (vertex A). The following is the mapping function from $w$-plane to $z$-plane [26]:

$$z(w) = M \int_{w_B}^{w} \frac{s - w_G}{\sqrt{s - w_B}\sqrt{s - w_E}\sqrt{s - w_F}\sqrt{s - w_H}} ds \qquad (2)$$

where $w_B, \ldots, w_H$ are the complex coordinates of the corresponding vertices in $w$-plane. $M$ is a complex constant. The complex coordinates of vertices C, D, E, F, G, H, and A in $z$-plane, namely, $z_C$, $z_D$, $z_E$, $z_F$, $z_G$, $z_H$, and $z_A$, respectively, are known, which are also the definite integrals of equation (2) when their upper limits are $w_C$, $w_D$, $w_E$, $w_F$, $w_G$, $w_H$, and $w_A$, respecitvely. These upper limits and $M$ are then determined based on these relationships. No analytical solution exists for these coordinates. Therefore, the complex coordinates of all vertices in $w$-plane are solved with an SC toolbox for MATLAB created by T. A. Driscoll [27, 28].

As discussed in the previous section, based on a forward SC mapping, the upper half-plane of $w$-plane can then be mapped onto a rectangle in another complex plane, namely, $t$-plane, as shown in Fig. 4d. The mapping function is

$$t(w) = N \int_{w_B}^{w} \frac{s - w_{E'}}{\sqrt{s - w_B}\sqrt{s - w_C}\sqrt{s - w_D}\sqrt{s - w_E}\sqrt{s - w_F}\sqrt{s - w_H}} ds \qquad (3)$$

$w_{E'}$ in above function is unknown and should be solved before the mapping is performed. As analyzed in the previous section, it should satisfy

$$\int_{w_E}^{w_F} \frac{s - w_{E'}}{\sqrt{s - w_B}\sqrt{s - w_C}\sqrt{s - w_D}\sqrt{s - w_E}\sqrt{s - w_F}\sqrt{s - w_H}} ds = 0 \qquad (4)$$

So $w_{E'}$ can be expressed as

$$w_{E'} = \frac{\int_{w_E}^{w_F} \frac{s}{\sqrt{s - w_B}\sqrt{s - w_C}\sqrt{s - w_D}\sqrt{s - w_E}\sqrt{s - w_F}\sqrt{s - w_H}} ds}{\int_{w_E}^{w_F} \frac{1}{\sqrt{s - w_B}\sqrt{s - w_C}\sqrt{s - w_D}\sqrt{s - w_E}\sqrt{s - w_F}\sqrt{s - w_H}} ds} \qquad (5)$$

After solving for $w_{E'}$, equation (3) can be solved numerically by MATLAB and the calculated electrostatic force is also plotted in Fig. 5a by the triangular dots. Compared with the theoretical results of [16], our strict results are obviously more consistent with the simulation results of [16]. This observation shows the effectiveness and accuracy of the conformal mapping method for multiple conductors.

A rotation micromirror based on the electrostatic repulsive actuator is developed in [16]. The theoretical and experimental performances (rotation angle versus applied voltage) achieved by He *et al*. are shown in Fig. 5b as the solid line and the triangular dots respectively. We recalculated the rotation angle based on our results of electrostatic force and the results are also shown as circular dots and dashed line, respectively in Fig. 5b. Compared with the theoretical values given in [16], our theoretical values based on the strict conformal mapping are considerably closer to the test values given in [16]. This also proves the statement proposed in [16], i.e., the infinite approximation used is one of the reasons for the discrepancy between the solid line and the triangular dots.

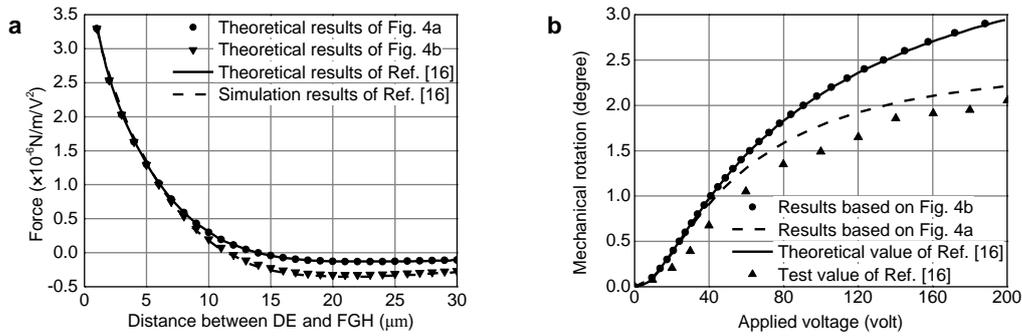

**Figure 5 | Analyzed and experimental results of Fig. 4. a**, The electrostatic force obtained in Ref. [16] and in this study. **b**, The rotation angle obtained in Ref. [16] and in this study.

Ref. [19] designed a T-shaped beam splitter, as indicated in Fig. 6a. Through carefully designing the refractive index distribution in the shaded region, the incoming beam through port 1 will be equally split into two beams, then smoothly pass through the 90° and −90° bend and finally exit through ports 2 and 3 respectively. In the literature the splitter was designed through solving Laplace's equations numerically. The results show that the maximum index appears near the corners C and H and is about 6.7 times the index in port 1. The minimum index appears near the midpoint of EF.

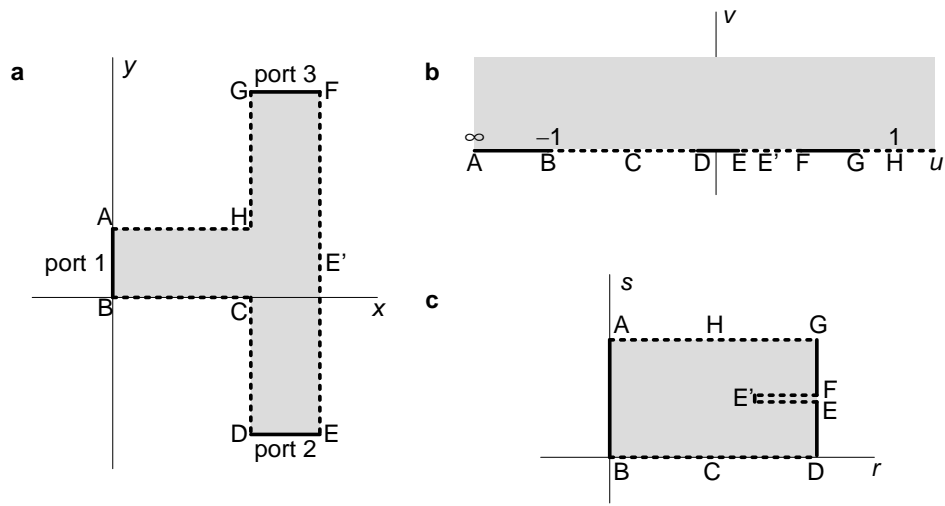

**Figure 6 | Conformal mapping processes for a beam splitter. a**, The structure of the beam splitter. The three solid lines represent three ports for input or output. **b**, The splitter is mapped onto the upper half-plane of *w*-plane. **c**, The final mapped pattern of the splitter.

As demonstrated in the calculation of the electrostatic actuator, we construct a

*z*-plane, as shown in Fig. 6a. Due to the design goal, the mapped splitter should look like Fig. 6c. Considering the symmetry, in *z*-plane the vertex E' should be just the midpoint of EF, as shown in Fig. 6a. An intermediate geometry, the upper half-plane of *w*-plane as shown in Fig. 6b, is still be used to construct the forward and inverse SC mapping onto the two complex planes, *z*-plane and *t*-plane. Similarly, the mapping function from *w*-plane to *z*-plane is

$$\frac{dz}{dw} = M \frac{\sqrt{w-w_C}\sqrt{w-w_H}}{\sqrt{w-w_B}\sqrt{w-w_D}\sqrt{w-w_E}\sqrt{w-w_F}\sqrt{w-w_G}} \tag{6}$$

The mapping function from *w*-plane to *t*-plane is

$$\frac{dt}{dw} = N \frac{w-w_{E'}}{\sqrt{w-w_B}\sqrt{w-w_D}\sqrt{w-w_E}\sqrt{w-w_F}\sqrt{w-w_G}} \tag{7}$$

The refractive indexs of the shaded region in *z*-plane, *w*-plane, and *t*-plane are denoted $n_z$, $n_w$, and $n_t$, respectively. They are determined by the following equations [17]

$$n_w = \left|\frac{dz}{dw}\right| n_z \tag{8}$$

$$n_w = \left|\frac{dt}{dw}\right| n_t \tag{9}$$

Assuming that the refractive index in *t*-plane is 1 (vacuum), the index in *z*-plane can be expressed as

$$n_z = \frac{\left|\frac{dt}{dw}\right|}{\left|\frac{dz}{dw}\right|} n_t = \left|\frac{w-w_{E'}}{\sqrt{w-w_C}\sqrt{w-w_H}}\right| \tag{10}$$

where the complex constant *M* and *N* are set as 1 for simplicity. Now the index $n_z$ to be solved is determined by a relatively simple formula. It is easy to know the maximum and minimum value of $n_z$, i.e., a maximum value of infinity at vertex C and H and a minimum value of 0 at vertex E', by using the numerator and denominator of the equation (10). Thus the position distribution of the index obtained here is consistent with [19]. As for the maximum value, we believe that the result in [19] is limited by the grid or step size of the numerical calculation.

**Conclusions**
In this paper a strict conformal mapping method for multiple terminals is proposed and described in detail by taking several typical situations as examples. Two application examples are calculated based on the method and the results are compared with those in the literatures, which validate the effectiveness and accuracy of the proposed method.

**Acknowledgements**

This work is supported by the National Natural Science Foundation of China (grant no. 11403029) and the Youth Innovation Promotion Association CAS (grant no. 2014346).

**Competing financial interests**

The authors declare no competing financial interests.